\documentclass[prd,twocolumn,amsmath,amssymb,aps,superscriptaddress,groupedaddress,nofootinbib,showpacs]{revtex4}
\usepackage{graphicx,float}
\usepackage[usenames, dvipsnames]{color}
\usepackage{tabularx}
\usepackage{siunitx}
\usepackage{xcolor}
\usepackage[
verbose,
colorlinks=true,
naturalnames=true,
linkcolor=blue,
citecolor=blue,
]{hyperref}

\begin{document}

\preprint{APS/123-QED}

\title{Hierarchical constraints on gravitational waves from horizonless compact objects}

\author{Rajrupa Mondal}
\affiliation{Maryland Center for Fundamental Physics, Department of Physics, University of Maryland, College
Park, MD 20742, USA}
\affiliation{Indian Institute of Science Education and Research, Kolkata, India}

\author{Julian Westerweck}
\affiliation{Institute for Gravitational Wave Astronomy and School of Physics and Astronomy, University of Birmingham, Edgbaston, Birmingham B15 2TT, United Kingdom}
\affiliation{Max-Planck-Institut f{\"u}r Gravitationsphysik (Albert-Einstein-Institut), \\ Callinstra{\ss}e 38, D-30167 Hannover, Germany}
\affiliation{Leibniz Universit{\"a}t Hannover, D-30167 Hannover, Germany}

\author{Yotam Sherf}
\affiliation{Max-Planck-Institut f{\"u}r Gravitationsphysik (Albert-Einstein-Institut), \\ Am M{\"u}hlenberg 1, D-14476 Potsdam, Germany}
\affiliation{Department of Physics, Ben-Gurion University, Beer-Sheva 84105, Israel}

\author{Collin D.\ Capano}
\affiliation{Department of Physics, Syracuse University, Syracuse, NY 13244, USA}
\affiliation{Physics Department, University of Massachusetts Dartmouth, North Dartmouth, MA 02747, USA}
\affiliation{Max-Planck-Institut f{\"u}r Gravitationsphysik (Albert-Einstein-Institut), \\ Callinstra{\ss}e 38, D-30167 Hannover, Germany}
\affiliation{Leibniz Universit{\"a}t Hannover, D-30167 Hannover, Germany}

\author{Ram Brustein}
\affiliation{Department of Physics, Ben-Gurion University, Beer-Sheva 84105, Israel}

\date{\today}

\begin{abstract}
We use the data of several promising gravitational wave observations to obtain increasingly stringent bounds on near-horizon deviations of their sources from the Kerr geometry.
A range of horizonless compact objects proposed as alternatives to black holes of general relativity would possess a modified gravitational wave emission after the merger.
Modelling these objects by introducing reflection of gravitational waves near the horizon, we can measure deviations from Kerr in terms of a single additional parameter, the location of the reflection.
We quote bounds on deviations for 5 events in addition to previous results obtained for GW150914.
Additionally, we improve upon previous results by hierarchically combining information from all analysed events, yielding a bound on deviations of less than $2.5 \times 10^{-26}$ meters above the horizon.
\end{abstract}

\maketitle

\section{Introduction}

Recent detections of the gravitational waves (GWs) emitted by merging compact objects provide a new way to probe strong-field gravitational physics and the nature of these compact objects.
While observations so far have been compatible with the signals expected from binary black holes (BHs) as predicted by general relativity (GR)~\cite{LIGOScientific:2016lio,Isi:2019aib, LIGOScientific:2020tif,Capano:2021etf,LIGOScientific:2021sio}, additional subtle features may still be present in the signals and reveal deviations from the GR-black hole hypothesis.
Such deviations are expected to appear at near-horizon scales on grounds of modifications of GR or alternative models for compact objects, typically inspired by considerations of quantum effects.

Black holes of GR are determined by their horizons, and the final object resulting from a binary black hole merger is expected to be approximated by the Kerr geometry with additional perturbations at sufficiently late times~\cite{Kerr:1963ud}.
A perturbed Kerr black hole emits gravitational waves in form of a quasi-normal mode (QNM) spectrum of short-lived damped sinusoids~\cite{Vishveshwara:1970zz,Teukolsky:1972my,Teukolsky:1973ha}.
The properties of this spectrum are directly affected by the presence of the event horizon.
If no horizon is present, the resulting quasi-normal mode spectrum instead consists of long-lived damped sinusoids~\cite{Starobinsky:1973aij,Vilenkin:1978uc,Maggio:2018ivz}.
The early part of the post-merger emission, however, remains unchanged, as the ringdown radiation originates at the photon sphere and the travel time to the horizon location prevents immediate contamination from the modified inner boundary~\cite{Cardoso:2016rao,Cardoso:2016oxy}.
Only at late times after the merger do the modified QNMs appear.
Therefore, observations of GWs with a primary inspiral-merger-ringdown emission compatible with black holes of GR still provide candidates for the search for additional late-time signals.

A number of alternative models have been proposed wherein no horizon is encountered by gravitational waves, thereby leading to the emission of modified QNMs.
Among these are boson stars and traversable wormholes, where GWs could pass through the interior, as well as gravastars and black holes with a near-horizon firewall, where physical reflection of GWs may occur~\cite{Mathur:2005zp,Guo:2017jmi,Brustein:2018ixz}.
These models share the property that their exterior is described by the Kerr geometry up to distances near the location of the horizon.
Their deviations from the Kerr description are then modelled approximately by introducing a reflective boundary near the horizon.
The location of this boundary determines the properties of the QNMs. Measuring their parameters allows us to infer the relative distance $\epsilon$ between the horizon and the proposed boundary.
Constraints on this distance serves as a probe of the Kerr description of the final object; an $\epsilon$ of zero indicates an object that is indistinguishable from a Kerr BH.

Electromagnetic observations have been used to place constraints on $\epsilon$ by considering a hard surface placed outside the horizon and relating a resulting electromagnetic emission to observed luminosities from black holes.
While strong bounds have been quoted in these studies~\cite{Doeleman:2012zc,EventHorizonTelescope:2019dse,EventHorizonTelescope:2019ggy,Zulianello:2020cmx,Lu:2017vdx}, they rely on a number of assumptions such as on the matter surrounding the black hole, with further discussion in~\cite{Lu:2017vdx,Carballo-Rubio:2018jzw}.

Gravitational wave detections have been used to constrain the presence of reflective near-horizon structure following two approaches, studying either the long-lived QNMs or so-called echoes.
GW echoes are a proposed signal of repeated damped pulses immediately following the ringdown, with the long-lived QNMs appearing at later times.
While several searches for echoes have been performed, no undisputed evidence for their presence in the data has been found, with a number of studies finding no significant evidence across many detections~\cite{Abedi:2016hgu,Ashton:2016xff,Conklin:2017lwb, Westerweck:2017hus,Nielsen:2018lkf,Uchikata:2019frs,LIGOScientific:2020tif,Ren:2021xbe,Wu:2023wfv}.
Indirect constraints on $\epsilon$ can be inferred from the absence of such a signal, but several uncertainties regarding modelling assumptions for echoes and a number of free parameters complicate placing strong bounds based on this signal~\cite{Cardoso:2019rvt}.
The effect of tidal heating on gravitational waveforms has been explored as an alternative avenue to probe the presence of horizons with next-generation detectors~\cite{Datta:2020gem,Mukherjee:2022wws}.

The late long-lived QNMs are determined by $\epsilon$ as the single additional parameter under a limited number of assumptions.
Due to its simpler modelling, this signal presents a promising candidate to place direct bounds on $\epsilon$, while presenting technical challenges due to its long-lived nature and expected small amplitude.
A study considering superradiant QNMs for the entire black hole population and targeting their effect on the gravitational wave background placed strong constraints under these assumptions~\cite{Barausse:2018vdb}.
Focussing instead on a single GW observation and searching for damped signals,~\cite{Westerweck:2021nue} provided the first direct individual constraints on deviations from the Kerr hypothesis in terms of $\epsilon$.

Here we expand the approach presented in~\cite{Westerweck:2021nue} to the most promising detections from the first 3 observing runs of detectors in the LIGO-Virgo-KAGRA network.
The same data-analysis techniques developed to accommodate the long-duration signal are applied here.
In addition to the individual measurements of $\epsilon$ for the selected events, we find a combined constraint based on the entirety of these events' data under the assumption that the corresponding compact objects share a common deviation from the Kerr geometry.

\section{Methods}

Black hole quasi-normal modes are found as the solution to the Teukolsky equation describing a perturbation on a Kerr black hole background with given boundary conditions~\cite{Teukolsky:1972my,Teukolsky:1973ha}.
At spatial infinity, the boundary conditions prescribe that gravitational radiation is purely out-going such that no waves are entering the system.
For a black hole, the horizon is a boundary of purely in-going waves.
The solution then consists of a spectrum of rapidly damped sinusoids, which are expected to approximate the post-merger emission of a binary merger remnant at times sufficiently late for this linear perturbative description to hold.

If instead reflective boundary conditions are introduced in the interior of the Kerr metric, the resulting quasi-normal modes are still a spectrum of damped sinusoids, but have far longer damping times.
The reflective boundary can represent either physical reflection or, in the case of reflection at the center, waves passing through the object~\cite{Starobinsky:1973aij,Vilenkin:1978uc}.
Here we approximate both cases by placing an effective reflective surface at a Boyer-Lindquist radius $r_{NH}$, a fractional distance $\epsilon$ above the horizon at $r_+$, such that $r_{NH} = r_+ (1 + \epsilon)$.
Solving the Teukolsky equation then allows to find the complex frequencies of the quasi-normal modes, $\omega = 2 \pi f + i \tau$, containing the central frequency $f$ and damping time $\tau$ of the damped sinusoid.

The initial ringdown of the horizonless object, however, is expected to remain unchanged compared to the black hole case.
As the ringdown radiation is associated with the photon sphere, the initial emission remains unaffected by the presence of an internal reflection due to the travel time to the point of reflection.
Only at later times would the post-merger signal begin to deviate from the black hole expectation.
Therefore, the observation of the black-hole ringdown in detected gravitational-wave signals does not rule out the presence of a modified quasi-normal mode spectrum at later times.

Additionally, more complicated structure may appear in the signal at times after the ringdown but before the long-lived modes.
This structure can take the form of repeated, damped pulses called echoes.
While there is significant uncertainty regarding the properties of the echo signal, it is expected to quickly damp to leave only the long-lived quasi-normal modes as the final signal.

In general, the signal depends on the reflectivity of the surface, which may be frequency-dependent. Here, we consider the reflection to be perfect. On theoretical grounds, it is argued that either full reflectivity or perfect absorption is expected for most alternative compact objects. The latter case would not lead to an additional late-time signal and be indistinguishable from the black hole case (see \cite{Mathur:2005zp,Guo:2017jmi} and Appendix of~\cite{Westerweck:2021nue}).

We additionally assume all energy falling in or extracted from the BH's rotation to be converted to the long-lived modes, ignoring potential earlier echoes. While this is an optimistic assumption for the considered signal, the uncertain description regarding an echo phase does not allow a reliable estimate of the energy likely contained therein.
A free parameter for the fraction of available energy converted to the long-lived modes would be degenerate with the measured amplitude parameter.
Therefore we determine the amplitude by the total available energy.

The resulting signal stabilises at late times to the quasi-normal modes of the compact object, taking the form of a long-lived damped sinusoid.
Following the signal model of~\cite{Westerweck:2021nue}, we focus on the mode expected to be dominant in the resulting spectrum, and describe the signal by
\begin{align}
    (h_{+} + i h_{\times}) (t) &= {}_{-2} S_{\ell m n} (\iota, \varphi, \chi) A e^{-t/\tau} e^{i (2 \pi f t + \phi)} \Theta(t-t_0),
    \label{ht}
\end{align}
with the amplitude $A$, frequency $f = \omega_R / 2 \pi$, damping time $\tau = \omega_I^{-1}$ and initial phase $\phi$ of the damped sinusoid, and the start time $t_0$ of the signal relative to the merger in the step-function $\Theta$.
The spin-weighted spheroidal harmonics ${}_{-2}S_{\ell m n}$ depend on the inclination $\iota$, azimuth angle $\varphi$, the black hole spin $\chi$, with the numbers $\ell,m,n = 2,2,0$ for the dominant mode.
We approximate the spheroidal harmonics by spin-weighted spherical harmonics, ${}_{-2} S_{\ell m n} \approx {}_{-2} Y_{\ell m}$ \cite{Berti:2007zu,Berti:2005gp}.

Following~\cite{Westerweck:2021nue}, the parameters determining the mode, $A, f$, and $\tau$, are given by
\begin{align}
    M \omega_R &= \frac{\chi}{ \left( 1 + \sqrt{1 - \chi^2} \right)} + \frac{\pi \sqrt{1 - \chi^2}}{ \left| \ln \epsilon \right| \left( 1 + \sqrt{1 - \chi^2} \right)}, \\
    \tau &= \dfrac{225M}{32\pi}\left(\dfrac{1+\sqrt{1-\chi^2}}{\sqrt{1-\chi^2}}\right)^6\dfrac{|\ln \epsilon|^7}{\left(\chi |\ln \epsilon|+\pi\sqrt{1-\chi^2}\right)^5}, \\
    A &= \dfrac{4}{\omega_R D_L  }\left(\dfrac{\Delta E}{\tau}\right)^{1/2},\\
    \Delta E &= E_{init}\left(1+\dfrac{ \chi^2}{8}\right).
\end{align}

Using this signal model and the gravitational-wave data surrounding detected events, we estimate the signal parameters through Bayesian inference methods.
As is common in gravitational-wave data analysis, we construct a likelihood function assuming the detector noise to be Gaussian. The likelihood for a signal with given parameters is calculated and the parameters' posterior probability distributions are estimated using the \texttt{PyCBC Inference} toolkit \cite{pycbcgithub, Biwer:2018osg}. Here, the parallel-tempered Markov-chain Monte Carlo sampler \texttt{emcee\_pt} is chosen to sample the parameter space \cite{ForemanMackey:2012ig,Vousden:2015}.

We employ two modifications to the standard Bayesian inference tools to accommodate the long-duration signal, heterodyning to reduce computational cost and treatment of boundary effects, as introduced in~\cite{Westerweck:2021nue}.

The heterodyning procedure makes use of the narrow frequency-domain representation of the signals we consider here.
We generate the signal templates in the time domain with the duration of the analysis window. The signal template is therefore a damped sinusoid starting and ending at the window edges. Generating the damped sinusoid directly in the frequency domain would instead lead to a self-overlapping signal due to the periodicity assumption of the Fourier transform. The late part of the signal would enter again at the beginning of the analysis window, leading to an unphysical template.
To reduce the computational cost, we can generate the signal in the time domain with a lower central frequency $f$ at a correspondingly lower and less expensive sampling frequency.
As the frequency representation of the signal is restricted to a narrow band around the central frequency, we can now shift the content of this band to the desired frequency without losing relevant signal content.
This procedure is contingent on choosing a frequency band for the generation of the signal that is sufficiently broad to capture all relevant signal content, so we adapt the width of the band depending on the width of the signal determined by its damping time.

We remove artefacts introduced by the interaction of the analysis window's boundaries and the whitening filter used in the matched-filtering process.
The periodicity assumption of the Fourier transform leads to a discontinuity of the data and waveform at the edges of the analysis window.
Applying the whitening filter then introduces "ringing" artefacts near the edges, resulting from the impulse response function of the filter.
For short signals with negligible contribution near the window boundaries, these are typically removed by introducing a tapering function reducing the amplitude to zero at the edges.
As the considered signal extends beyond the analysis window, such a tapering is undesirable as it would also interact with the signal waveform.
Instead, we use the following procedure.

We are interested in analysing data from time $t_1$ to time $t_2$. We begin by selecting data from  $t_1-\Delta t$ to $t_2+\Delta t$. Similarly, we generate the template as a timeseries starting at $t_1-\Delta t$ with a duration of $t_2-t_1+2\Delta t$, adjusting the initial amplitude of the damped sinusoid by a factor $\exp [\Delta t / \tau]$. Both the data and template are Fourier transformed to the frequency domain where the whitening filter is applied. Then the whitened data and template are transformed back to the time domain, where we remove the additional data containing the introduced artefacts, i.e.\ $(t_1-\Delta t,t_1)$ and $(t_2,t_2+\Delta t)$. $\Delta t$ is chosen such that the artefacts are negligible outside of the removed data.
Finally, the remaining whitened data and template are transformed back to the frequency domain where the likelihood is calculated from their inner product.

We restrict the analysis to the most promising events due to its computational cost.
Events are selected based on an estimate of their expected optimal signal-to-noise ratio (SNR) for the modified quasi-normal mode signal.
We calculate the optimal SNR by generating a simulated signal and calculating the matched-filter SNR with a template identical to the signal, using a power-spectral density estimated from the data before the detected real gravitational-wave signal.
In this, we use the modified inference procedure described above.
Where possible, the values for the parameters of the simulated signals are either chosen to be the maximum-likelihood values from the posteriors of~\cite{Nitz:2021zwj} or derived from these.
The mass and spin of the merger remnant and the energy radiated in the primary gravitational-wave emission are calculated from the measured component masses and spins via fitting formulae to numerical relativity results \cite{Hofmann:2016yih, Tichy:2008du, lalsuite}.
For the sole additional parameter not determined through this, $\epsilon$, we choose a fixed value from within our prior range, $\epsilon = 10^{-15}$.
While this does not address more subtle effects, such as the dependency of the SNR on the combination of the signal frequency and corresponding PSD, it provides an estimate to select a few promising events.
We choose an integration time of 128 seconds for this estimate.
The events selected for our analysis are listed in Table~\ref{tab:results}, together with the estimated SNRs.
The first five events have the highest estimated SNRs, while GW200224 is chosen due to its parameters being similar to those of GW150914, for which a similar analysis was first performed~\cite{Westerweck:2021nue}.
We have omitted GW191109\_010717 and
GW190424\_180648 from our analysis despite their similarly high predicted SNRs due to data quality issues, such as not enough data prior to the event to estimate the PSD.

We can combine information from multiple detections to find a combined posterior distribution for parameters that are identical between the individual events.
Here we assume that the relative separation between the horizon and the reflective region, $\epsilon$, is the same for the remnant objects in the selected events.
We can then analyse each event separately and combine the resulting $\epsilon$-posteriors as follows.
Let $\mathcal{P}_{i}(\epsilon)$ be the prior on $\epsilon$ for event $i$, while the datasets $x_i$ of the events are independent. We use Bayes' Theorem,
\begin{align}
\mathcal{P}(\epsilon|x_{i}) = \frac{\mathcal{P}(x_{i}|\epsilon) \mathcal{P}_{i}(\epsilon)}{\mathcal{P}(x_i)} \propto \mathcal{P}(x_{i}|\epsilon) \mathcal{P}_{i}(\epsilon).
\end{align}

To find a combined posterior for, e.g., two events with data $x_1,x_2$, we use the posterior of the first event as the prior for the second event, giving
\begin{align}
\mathcal{P}(\epsilon|x_2,x_1) &= \frac{\mathcal{P}(x_2|\epsilon, x_1) \mathcal{P}(\epsilon|x_1)}{\mathcal{P}(x_2|x_1)} = \frac{\mathcal{P}(x_2|\epsilon) \mathcal{P}(\epsilon|x_1)}{\mathcal{P}(x_2) \mathcal{P}(x_1)}\\
&= \frac{\mathcal{P}(\epsilon|x_2) \mathcal{P}(x_2)}{\mathcal{P}_2(\epsilon)} \frac{\mathcal{P}(\epsilon|x_1)}{\mathcal{P}(x_2) \mathcal{P}(x_1)}\\
&= \frac{\mathcal{P}(\epsilon|x_2)}{\mathcal{P}_2(\epsilon)} \frac{\mathcal{P}(\epsilon|x_1)}{\mathcal{P}_1(\epsilon) \mathcal{P}(x_1)} \mathcal{P}_1(\epsilon)\\
&\propto \frac{\mathcal{P}(\epsilon|x_{2})}{\mathcal{P}_{2}(\epsilon)}\frac{\mathcal{P}(\epsilon|x_{1})}{\mathcal{P}_{1}(\epsilon)}\mathcal{P}_{1}(\epsilon),
\end{align}
where we used Bayes' theorem, the independence of $x_1$ and $x_2$, and in the last line multiplied by 1.

Generalizing to $N$ events, we recursively use the posterior of the previous $N-1$ events as the prior for the $N$-th event and obtain the relation
\begin{align}
\mathcal{P}(\epsilon|x_N,x_{N-1},...,x_1)\propto \prod_{i=1}^{N} \frac{\mathcal{P}(\epsilon|x_{i})}{\mathcal{P}_{i}(\epsilon)}\mathcal{P}_{1}(\epsilon).
\end{align}
When the priors of all datasets are assumed to be the same, $\mathcal{P}_i(\epsilon) = \mathcal{P}(\epsilon) \forall i$, the relation simplifies to
\begin{align}
\mathcal{P}(\epsilon|x_N,x_{N-1},...,x_1)\propto \frac{\prod_{i=1}^{N} \mathcal{P}(\epsilon|x_{i})}{\mathcal{P}(\epsilon)^{N-1}}. \label{eq:hierarchical_posterior}
\end{align}
We apply this relation as we use the same prior on $\log_{10} \epsilon$ for each event.
As the final posterior distribution will be normalised, the proportionality in this relation is sufficient for our application.

In addition to $\epsilon$, we treat all parameters of the final black hole as free and vary them during parameter estimation.
The priors for these are chosen to be the posteriors from the analysis of the full inspiral-merger-ringdown (IMR) signal given in~\cite{Nitz:2021zwj}. The mass and spin of the final black hole are again calculated from the measurement of the binary parameters using fits to numerical relativity results \cite{Hofmann:2016yih,Tichy:2008du,lalsuite}.
We choose a prior linear in $\log_{10} \epsilon$ as was used in \cite{Westerweck:2021nue}, as the modified QNM signal's parameters depend on $\epsilon$ logarithmically. The range for this covers many order of magnitude, with $\log_{10} \epsilon \in [-45,-2]$.
For GW190929, we impose an additional constraint on the mode's central frequency, $f \notin [55,65]$ Hz, to avoid contamination from PSD fluctuations due to the power grid at $60$ Hz.

We start the analysis for the modified QNM signal at 2 seconds after the merger, to prevent contamination from potentially present earlier post-ringdown signals not captured by our signal model. Analysing long stretches of data is susceptible to confusing slowly varying, non-Gaussian noise features with signals, leading to weaker constraints on $\epsilon$. 
We therefore use progressively longer segments of data to find the optimum duration for the analysis, i.e.\ the longest segment that does not lead to weaker bounds compared to the previous choice.
Following this, we use 256 seconds of data for GW190521\_030229, and 128 seconds of data for all other events.
This restriction effectively prevents incorrectly detecting noise fluctuations at the cost of lower sensitivity to the signal.
The PSD is estimated from 512 seconds of data prior to the IMR signal.

\section{Results}
For GW150914, we use the results found in \cite{Westerweck:2021nue} instead of repeating the analysis for this event, as the setup for the analysis of an individual event is virtually unchanged.
Our analysis is now applied separately to the additional 5 events listed in Table~\ref{tab:results}.

\begin{figure}
    \centering
    \includegraphics[width=\linewidth]{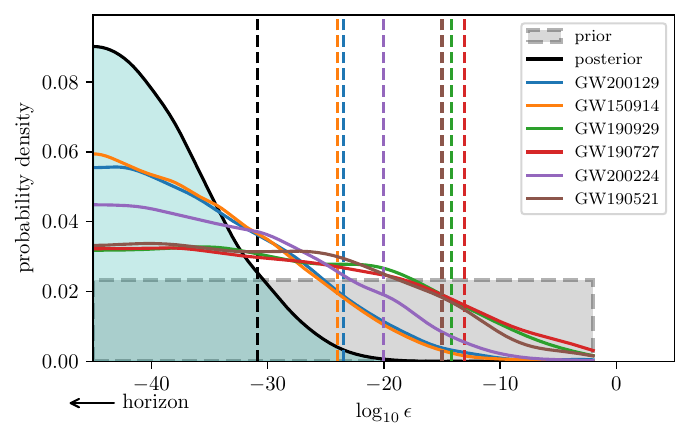}
    \caption{Posteriors of $\log_{10} \epsilon$ obtained from post-merger analysis of the listed events.
    The corresponding 90\% credible upper bounds are marked by dashed lines. The combined posterior is marked by the solid black contour and shaded region.
    The dashed grey shaded region shows the uniform prior of $\log_{10} \epsilon$.}
    \label{fig:epsilon_posteriors}
\end{figure}

\begin{figure}
    \centering
    \includegraphics[width=\linewidth]{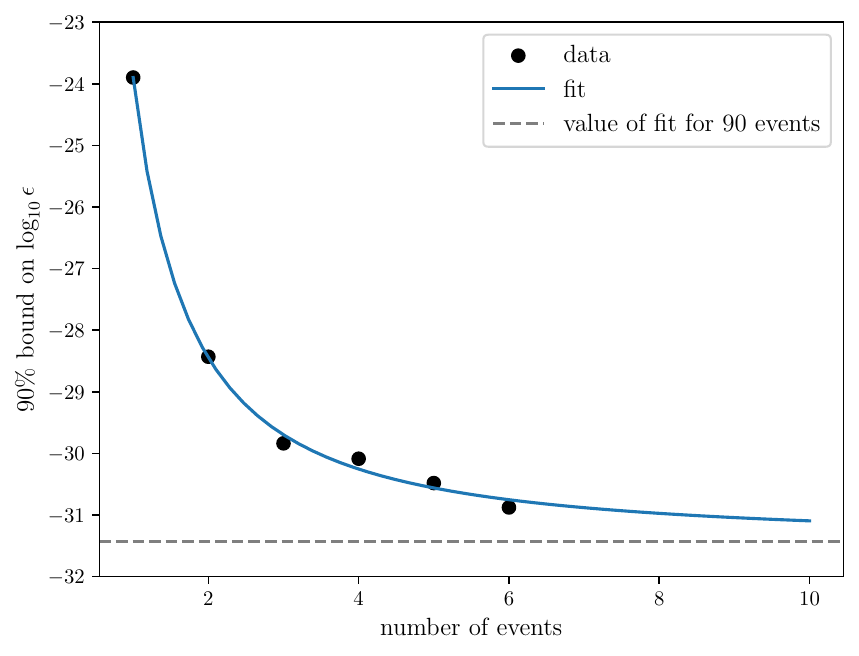}
    \caption{The one-sided 90\% credible upper bound of $\log_{10} \epsilon$ obtained by combining data from multiple events as a function of the number of events included. The fit is given by $y=7.55x^{-1.33}-31.45$ and the curve asymptotes to $y=-31.43$.}
    \label{fig:epsilon_evolution}
\end{figure}

The results for the individual and combined posteriors for $\log_{10} \epsilon$ are shown in Figures~\ref{fig:epsilon_posteriors}, \ref{fig:epsilon_evolution}, and Table~\ref{tab:results}.
Figure~\ref{fig:epsilon_posteriors} shows the posterior distribution obtained for each event, with the one-sided 90\% upper credible bound marked by the dashed lines and listed in Table~\ref{tab:results}. While GW150914 still yields the strongest individual bound at $\log_{10} \epsilon = -24$, combining the posteriors into a hierarchical result improves this to $\log_{10} \epsilon = -30.9$ for the 6 events analysed.
To calculate the product of posteriors in Equation~\ref{eq:hierarchical_posterior}, we convert the posterior samples to density functions.
A kernel-density estimation (KDE) using Gaussian kernels is used to approximate the posterior distributions based on the samples.
We use the \texttt{gaussian\_kde} function provided in the \texttt{scipy} software package \cite{Virtanen:2019joe}.
The KDE forces the probability density to be zero at the chosen boundaries and results in a smooth transition to this value.
This is unsuitable if the actual distribution peaks at or near the boundaries, as is the case for our results on $\log_{10} \epsilon$. We correct this by artificially extending the range of the KDE beyond the lower prior bound and mirroring the samples at this boundary, allowing the KDE to cover the region around the boundary. The desired probability density is then found by removing the artificially introduced half of the symmetric KDE below the actual prior boundary.

\begin{table}
\centering
\begin{tabular}[t]{c c c c c} 
\hline \hline
Event & optimal SNR & $\log_{10} \epsilon$ & $\log_{10} \Delta r$ & $\log_{10} \Delta s$\\
\hline
GW200129\_065458 & $6.67$ & $-23.44$ & $-18.25$ & $-6.59$ \\
GW150914\_095045 & $5.10$ & $-24.0$ & $-18.8$ & $-6.83$ \\
GW190929\_012149 & $4.06$ & $-14.18$ & $-8.77$ & $-1.37$ \\
GW190727\_060333 & $3.99$ & $-13.08$ & $-7.87$ & $-0.99$ \\
GW200224\_222234 & $2.61$ & $-20.01$ & $-14.78$ & $-4.83$ \\
GW190521\_030229 & $1.57$ & $-14.99$ & $-9.42$ & $-1.57$ \\
\hline \hline
\end{tabular}

\caption{The five most promising events in terms of the optimal SNR for the proposed post-merger signal are listed in the first column in addition to a candidate chosen for similarity to GW150914. The second column shows the optimal SNR calculated for the maximum-likelihood parameters from~\cite{Nitz:2021zwj}, with $\epsilon = 10^{-15}$ and an integration time of 128 seconds.
The third column shows the upper one-sided 90\% credible interval values of $\text{log}_{10}\epsilon$ for each individual event.
The fourth and fifth columns show the corresponding values for the coordinate distance and proper distance, $\log_{10} \Delta r$ and $\log_{10} \Delta s$, measured in meters, respectively.
}
\label{tab:results}
\end{table}

We use the individual results to predict the bounds that could be achieved by analysing a larger number of events.
For this, we calculate the bound found when increasing the number of individual posteriors included in the combined result.
As we expect further events to have less potential SNR in the modified QNM signal, we order the events by increasing upper bound to capture this trend.
Figure~\ref{fig:epsilon_evolution} shows the evolution of the combined upper bound depending on the number of posteriors included.
Fitting a curve of the form $y = ax^b + c$ for $a,b,c$ to the obtained bounds as a function of the number of included events, we extrapolate the bound achievable by including 90 events to be $\log_{10} \epsilon = -31.4$.


One may expect the absolute separation distance of the reflective boundary and the horizon to be the parameter shared between the individual events instead of the quantity $\epsilon$ which depends on the compact object's mass.
This could be the case if an additional length scale is introduced, motivated for example by quantum effects near the horizon.
Therefore, we also convert our results to show the posterior distribution for a measure of this separation.
Calculating the coordinate distance between the horizon and the reflective boundary in Boyer-Lindquist coordinates in the equatorial plane, we find the relation
\begin{align}
\Delta r &= \epsilon m_f(1+\sqrt{1-a_f^2}),
\label{eqn:coord_distance}
\end{align}
where $m_f,a_f$ are the final mass and spin of the remnant object, respectively, and $\Delta r$ will be measured in meters in the following.
This is the separation as seen by a far-away observer, while for the corresponding proper distance the relation is
\begin{align}
 \Delta s &= m_f \sqrt{2 \epsilon \left( 1-a_f^2 + \sqrt{1-a_f^2} \right)} \nonumber \\
 &\quad + m_f \: \text{arccosh} \left( 1 + \epsilon + \frac{\epsilon}{\sqrt{1 - a_f^2}} \right)
\end{align}

\begin{figure}
    \centering
    \includegraphics[width=\linewidth]{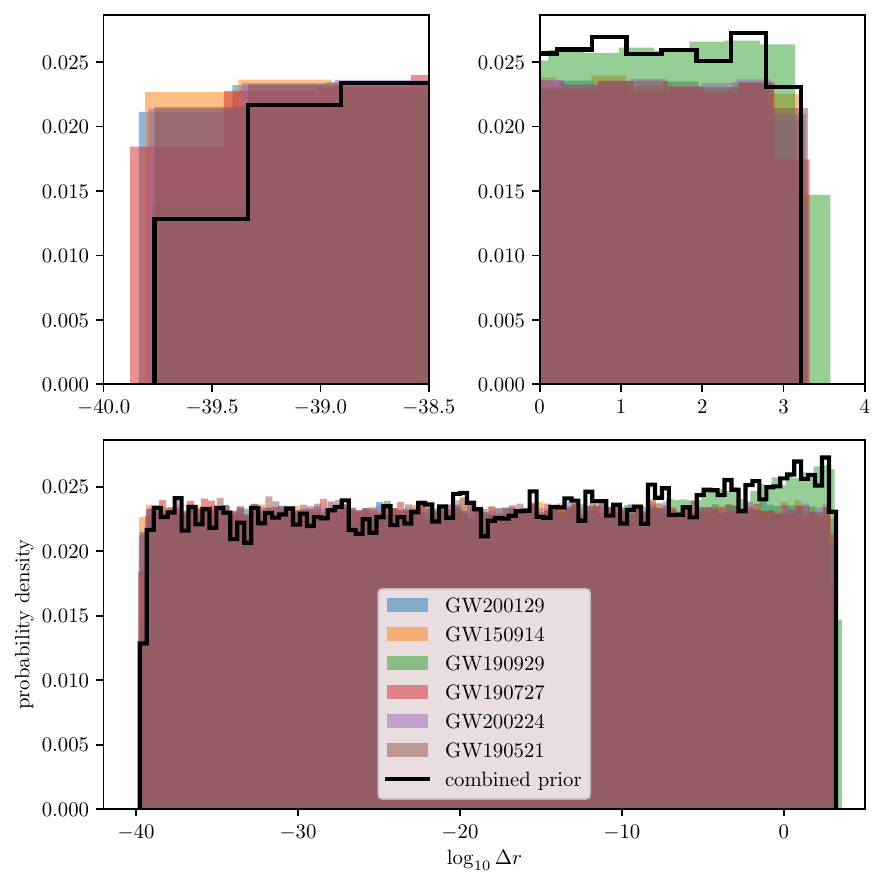}
    \caption{The priors on coordinate distance, $\log_{10} \Delta r$. The top panels show a zoom-in of the lower and upper bounds of the priors.
    Individual priors are shaded, while the combined prior is shown by the black contour. The priors are approximately uniform, deviating from this only at the boundaries of the allowed range.
    }
    \label{fig:fig_coord_dist_priors}
\end{figure}

We find the posterior for $\log_{10} \Delta r$ by applying Eq.~\eqref{eqn:coord_distance}  to random draws from the joint posterior on $\log_{10} \epsilon$, $m_f$, and $a_f$.
The prior for $\log_{10} \Delta r$ is calculated for each event in the same way from the respective parameters' priors. The priors are slightly different due to the different IMR posteriors used as priors for our analysis. While the $\log_{10} \Delta r$ priors are therefore not perfectly uniform, they approach a uniform distribution with slightly different ranges for the individual events. The resulting priors are illustrated in Figure~\ref{fig:fig_coord_dist_priors}.
A small deviation from uniformity at the upper bound of GW190929's prior is due to the additional constraint on the mode frequency imposed for this detection.
To find combined posteriors, we use as the prior range for $\log_{10} \Delta r$ the intersection of the individual prior ranges, and divide each event's posterior by its respective prior.

\begin{figure}
    \centering
    \includegraphics[width=\linewidth]{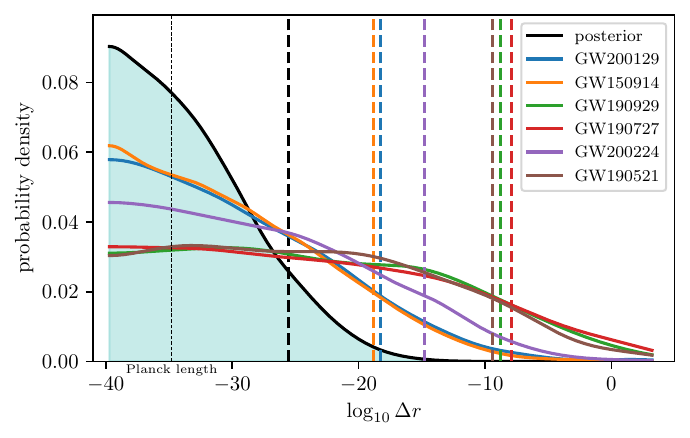}
    \caption{Posteriors of $\log_{10} \Delta r$ obtained from post-merger analysis of the listed events.
    $\Delta r$ is measured in meters.
    The corresponding 90\% credible upper bounds are marked by dashed lines. The combined posterior is marked by the black contour and shaded region.}
    \label{fig:coord_dist_posteriors}
\end{figure}

\begin{figure}
    \centering
    \includegraphics[width=\linewidth]{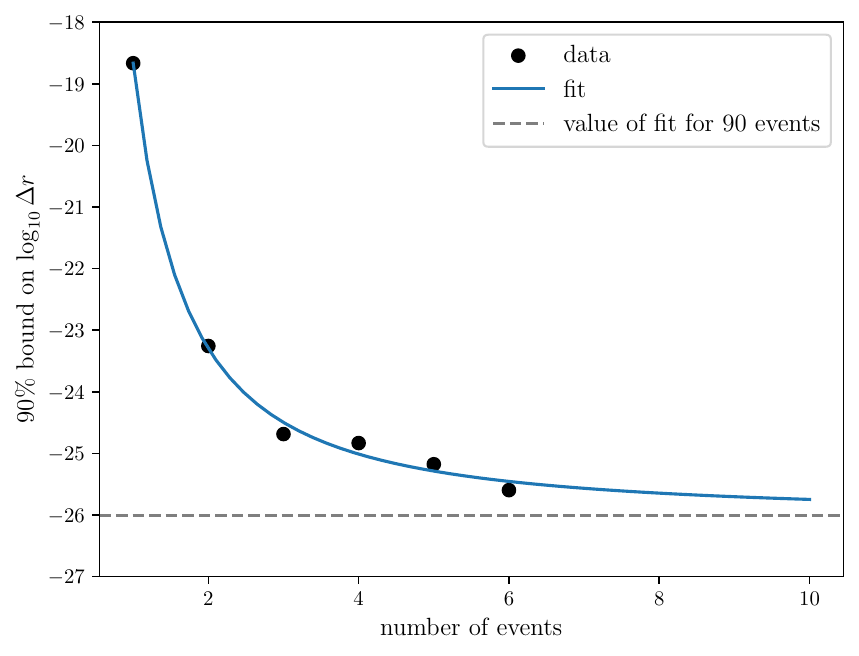}
    \caption{The one-sided 90\% credible upper bound of $\log_{10} \Delta r$ obtained by combining data from multiple events as a function of the number of events included. The fit is given by $y=7.35x^{-1.43}-26.02$ and the curve asymptotes to $y=-26.01$.}
    \label{fig:coord_dist_evolution}
\end{figure}

The results for $\log_{10} \Delta r$ are presented similarly to those for $\log_{10} \epsilon$, in Figures~\ref{fig:coord_dist_posteriors}, \ref{fig:coord_dist_evolution}, and Table~\ref{tab:results}.
Figure~\ref{fig:coord_dist_posteriors} shows the posterior distribution obtained for each event, with the one-sided 90\% upper credible bound marked by the dashed lines and listed in Table~\ref{tab:results}.
The table also lists the corresponding values for the proper distance $\Delta s$.
Again, GW150914 yields the strongest individual bound at $\log_{10} \Delta r \approx -18.8$.
The hierarchically combined posterior results in a 90\% upper bound of $\log_{10} \Delta r \approx -25.6$ for the six events analysed.

Figure~\ref{fig:coord_dist_evolution} shows the upper bound as a function of the number of events included in the hierarchical analysis.
Using the same method to extrapolate the results for the inclusion of further events, we find a predicted bound for 90 events of $\log \Delta r = -26$.
 
\section{Conclusions}
We have analysed a number of promising gravitational wave detections for evidence of modified quasi-normal mode signals as predicted for horizonless compact objects.
Using methods adapted to the long duration of the expected signals, we place bounds on their determining parameters for 5 detections in addition to GW159014, for which a similar study had been performed previously.
Combining information from all analysed events allows us to place a joint constraint on the relative location of deviations from the Kerr metric of $\log_{10} \epsilon = -30.9$, improving upon previous results from studying only a single event.
An additional bound on the coordinate distance between the horizon and the deviation from the Kerr geometry is derived from our results, giving $\log_{10} \Delta r = -25.6$, calculated in Boyer-Lindquist coordinates.

Our constraints are primarily limited by two factors, the detector sensitivity for available detections and the maximum duration of data that can be accommodated by our analysis.
Both are expected to improve in the near future:
The detectors are undergoing upgrades in sensitivity for the current observation run. Data-analysis methods to address individual non-Gaussian line features in the PSD can help extending the analysis to larger amounts of data.

Considering these improvements, we may expect to place bounds on the deviation on the order of the Planck length in the future.
If the model leading to the described signal is accurate, this would effectively rule out horizonless objects consisting of non-exotic matter as candidates for the analysed events' final objects.

However, the assumptions and simplifications of this model must be addressed to produce more reliable physical conclusions.
In particular these are ignoring the early post-ringdown phase of the signal, assuming all in-going energy to be converted to the late long-lived modes, and perfect effective reflectivity.

Incorporating these requires accurate modelling of the complete signal expected for alternative compact objects, stressing the importance of this avenue of research.
Until such models are available, or to cover uncertainties in their description, a first step could be a joint approach adding weakly modelled searches to our analysis.
BayesWave~\cite{Cornish:2014kda} and cWB~\cite{Drago:2020kic}could allow to capture emissions immediately after the ringdown and estimate the model parameters jointly with our modelled late-time signal search.

We plan to apply these improvements in future studies, aiming to produce bounds on alternative models for compact objects with increased robustness under relaxed modelling assumption and higher sensitivity.
Reaching Planck-level bounds can effectively rule out individual proposals and help focus further theoretical modelling efforts on the most promising candidates.
Combined with further precision measurements and theoretical development, this approach will contribute to determining the fundamental nature of astrophysical black holes.
 
\section{Acknowledgements}
Computations were performed on Unity, a collaborative, multi-institutional high-performance computing cluster managed by UMass Amherst Research Computing and Data.

Further calculations were performed on the Atlas computer cluster of the Max Planck-Institute for Gravitational Physics (Albert Einstein Institute) Hannover.

This research has made use of data obtained from the Gravitational Wave Open Science Center (https://www.gw-openscience.org/ ), a service of LIGO Laboratory, the LIGO Scientific Collaboration and the Virgo Collaboration. LIGO Laboratory and Advanced LIGO are funded by the United States National Science Foundation (NSF) who also gratefully acknowledge the Science and Technology Facilities Council (STFC) of the United Kingdom, the Max-Planck-Society (MPS), and the State of Niedersachsen/Germany for support of the construction of Advanced LIGO and construction and operation of the GEO600 detector. Additional support for Advanced LIGO was provided by the Australian Research Council. Virgo is funded, through the European Gravitational Observatory (EGO), by the French Centre National de Recherche Scientifique (CNRS), the Italian Istituto Nazionale di Fisica Nucleare (INFN) and the Dutch Nikhef, with contributions by institutions from Belgium, Germany, Greece, Hungary, Ireland, Japan, Monaco, Poland, Portugal, Spain. The research of Y.S.\ was supported by a Minerva Fellowship of the Minerva Stiftung Gesellschaft f{\"u}r die Forschung
mbH.

\bibliography{main_submission.bib}

@article{Nitz:2021zwj,
    author = {Nitz, Alexander H. and Kumar, Sumit and Wang, Yi-Fan and Kastha, Shilpa and Wu, Shichao and Sch\"afer, Marlin and Dhurkunde, Rahul and Capano, Collin D.},
    title = "{4-OGC: Catalog of Gravitational Waves from Compact Binary Mergers}",
    eprint = "2112.06878",
    archivePrefix = "arXiv",
    primaryClass = "astro-ph.HE",
    doi = "10.3847/1538-4357/aca591",
    journal = "Astrophys. J.",
    volume = "946",
    number = "2",
    pages = "59",
    year = "2023"
}

@article{Hofmann:2016yih,
      author         = "Hofmann, Fabian and Barausse, Enrico and Rezzolla,
                        Luciano",
      title          = "{The final spin from binary black holes in quasi-circular
                        orbits}",
      journal        = "Astrophys. J.",
      volume         = "825",
      year           = "2016",
      number         = "2",
      pages          = "L19",
      doi            = "10.3847/2041-8205/825/2/L19",
      eprint         = "1605.01938",
      archivePrefix  = "arXiv",
      primaryClass   = "gr-qc",
      SLACcitation   = "%%CITATION = ARXIV:1605.01938;%%"
}

@article{Tichy:2008du,
      author         = "Tichy, Wolfgang and Marronetti, Pedro",
      title          = "{The Final mass and spin of black hole mergers}",
      journal        = "Phys. Rev.",
      volume         = "D78",
      year           = "2008",
      pages          = "081501",
      doi            = "10.1103/PhysRevD.78.081501",
      eprint         = "0807.2985",
      archivePrefix  = "arXiv",
      primaryClass   = "gr-qc",
      SLACcitation   = "%%CITATION = ARXIV:0807.2985;%%"
}

@Misc{lalsuite,
  author       = {{LIGO Scientific Collaboration}},
  title        = {{LIGO} {A}lgorithm {L}ibrary - {LALS}uite},
  howpublished = {https://github.com/lscsoft/lalsuite},
  doi          = {10.7935/GT1W-FZ16},
}

@article{Kerr:1963ud,
    author = "Kerr, Roy P.",
    title = "{Gravitational field of a spinning mass as an example of algebraically special metrics}",
    doi = "10.1103/PhysRevLett.11.237",
    journal = "Phys. Rev. Lett.",
    volume = "11",
    pages = "237--238",
    year = "1963"
}

@article{LIGOScientific:2016lio,
    author = "Abbott, B. P. and others",
    collaboration = "LIGO Scientific, Virgo",
    title = "{Tests of general relativity with GW150914}",
    eprint = "1602.03841",
    archivePrefix = "arXiv",
    primaryClass = "gr-qc",
    reportNumber = "LIGO-P1500213",
    doi = "10.1103/PhysRevLett.116.221101",
    journal = "Phys. Rev. Lett.",
    volume = "116",
    number = "22",
    pages = "221101",
    year = "2016",
    note = "[Erratum: Phys.Rev.Lett. 121, 129902 (2018)]"
}

@article{Isi:2019aib,
      author         = "Isi, Maximiliano and Giesler, Matthew and Farr, Will M.
                        and Scheel, Mark A. and Teukolsky, Saul A.",
      title          = "{Testing the no-hair theorem with GW150914}",
      journal        = "Phys. Rev. Lett.",
      volume         = "123",
      year           = "2019",
      number         = "11",
      pages          = "111102",
      doi            = "10.1103/PhysRevLett.123.111102,
                        10.1103/physrevlett.123.111102",
      eprint         = "1905.00869",
      archivePrefix  = "arXiv",
      primaryClass   = "gr-qc",
      reportNumber   = "LIGO-P1900135",
      SLACcitation   = "%%CITATION = ARXIV:1905.00869;%%"
}

@article{LIGOScientific:2020tif,
    author = "Abbott, R. and others",
    collaboration = "LIGO Scientific, Virgo",
    title = "{Tests of general relativity with binary black holes from the second LIGO-Virgo gravitational-wave transient catalog}",
    eprint = "2010.14529",
    archivePrefix = "arXiv",
    primaryClass = "gr-qc",
    reportNumber = "LIGO-P2000091",
    doi = "10.1103/PhysRevD.103.122002",
    journal = "Phys. Rev. D",
    volume = "103",
    number = "12",
    pages = "122002",
    year = "2021"
}

@article{Capano:2021etf,
    author = "Capano, Collin D. and Cabero, Miriam and Westerweck, Julian and Abedi, Jahed and Kastha, Shilpa and Nitz, Alexander H. and Wang, Yi-Fan and Nielsen, Alex B. and Krishnan, Badri",
    title = "{Multimode Quasinormal Spectrum from a Perturbed Black Hole}",
    eprint = "2105.05238",
    archivePrefix = "arXiv",
    primaryClass = "gr-qc",
    doi = "10.1103/PhysRevLett.131.221402",
    journal = "Phys. Rev. Lett.",
    volume = "131",
    number = "22",
    pages = "221402",
    year = "2023"
}

@article{Brustein:2018ixz,
    author = "Brustein, Ram and Medved, A. J. M. and Yagi, K.",
    title = "{Lower limit on the entropy of black holes as inferred from gravitational wave observations}",
    eprint = "1811.12283",
    archivePrefix = "arXiv",
    primaryClass = "gr-qc",
    reportNumber = "CERN-TH-2018-261",
    doi = "10.1103/PhysRevD.100.104009",
    journal = "Phys. Rev. D",
    volume = "100",
    number = "10",
    pages = "104009",
    year = "2019"
}

@article{Vilenkin:1978uc,
    author = "Vilenkin, A.",
    title = "{Exponential Amplification of Waves in the Gravitational Field of Ultrarelativistic Rotating Body}",
    doi = "10.1016/0370-2693(78)90027-8",
    journal = "Phys. Lett. B",
    volume = "78",
    pages = "301--303",
    year = "1978"
}

@article{Starobinsky:1973aij,
    author = "Starobinsky, A. A.",
    title = "{Amplification of waves reflected from a rotating ''black hole''.}",
    journal = "Sov. Phys. JETP",
    volume = "37",
    number = "1",
    pages = "28--32",
    year = "1973"
}

@article{Maggio:2018ivz,
    author = "Maggio, Elisa and Cardoso, Vitor and Dolan, Sam R. and Pani, Paolo",
    title = "{Ergoregion instability of exotic compact objects: electromagnetic and gravitational perturbations and the role of absorption}",
    eprint = "1807.08840",
    archivePrefix = "arXiv",
    primaryClass = "gr-qc",
    doi = "10.1103/PhysRevD.99.064007",
    journal = "Phys. Rev. D",
    volume = "99",
    number = "6",
    pages = "064007",
    year = "2019"
}

@Article{Cardoso:2016rao,
  author        = {Cardoso, Vitor and Franzin, Edgardo and Pani, Paolo},
  title         = {{Is the gravitational-wave ringdown a probe of the event horizon?}},
  journal       = {Phys. Rev. Lett.},
  year          = {2016},
  volume        = {116},
  number        = {17},
  pages         = {171101},
  note          = {[Erratum: Phys.Rev.Lett. 117, 089902 (2016)]},
  archiveprefix = {arXiv},
  doi           = {10.1103/PhysRevLett.116.171101},
  eprint        = {1602.07309},
  primaryclass  = {gr-qc},
}

@Article{Cardoso:2016oxy,
  author        = {Cardoso, Vitor and Hopper, Seth and Macedo, Caio F. B. and Palenzuela, Carlos and Pani, Paolo},
  title         = {{Gravitational-wave signatures of exotic compact objects and of quantum corrections at the horizon scale}},
  journal       = {Phys. Rev. D},
  year          = {2016},
  volume        = {94},
  number        = {8},
  pages         = {084031},
  archiveprefix = {arXiv},
  doi           = {10.1103/PhysRevD.94.084031},
  eprint        = {1608.08637},
  primaryclass  = {gr-qc},
}

@Article{Abedi:2016hgu,
  author        = {Abedi, Jahed and Dykaar, Hannah and Afshordi, Niayesh},
  title         = {{Echoes from the Abyss: Tentative evidence for Planck-scale structure at black hole horizons}},
  journal       = {Phys. Rev. D},
  year          = {2017},
  volume        = {96},
  number        = {8},
  pages         = {082004},
  archiveprefix = {arXiv},
  doi           = {10.1103/PhysRevD.96.082004},
  eprint        = {1612.00266},
  primaryclass  = {gr-qc},
}

@article{Ashton:2016xff,
    author = "Ashton, Gregory and Birnholtz, Ofek and Cabero, Miriam and Capano, Collin and Dent, Thomas and Krishnan, Badri and Meadors, Grant David and Nielsen, Alex B. and Nitz, Alex and Westerweck, Julian",
    title = "{Comments on: ''Echoes from the abyss: Evidence for Planck-scale structure at black hole horizons''}",
    eprint = "1612.05625",
    archivePrefix = "arXiv",
    primaryClass = "gr-qc",
    month = "12",
    year = "2016"
}

@Article{Conklin:2017lwb,
  author        = {Conklin, Randy S. and Holdom, Bob and Ren, Jing},
  title         = {{Gravitational wave echoes through new windows}},
  journal       = {Phys. Rev. D},
  year          = {2018},
  volume        = {98},
  number        = {4},
  pages         = {044021},
  archiveprefix = {arXiv},
  doi           = {10.1103/PhysRevD.98.044021},
  eprint        = {1712.06517},
  primaryclass  = {gr-qc},
}

@article{Westerweck:2017hus,
    author = "Westerweck, Julian and Nielsen, Alex and Fischer-Birnholtz, Ofek and Cabero, Miriam and Capano, Collin and Dent, Thomas and Krishnan, Badri and Meadors, Grant and Nitz, Alexander H.",
    title = "{Low significance of evidence for black hole echoes in gravitational wave data}",
    eprint = "1712.09966",
    archivePrefix = "arXiv",
    primaryClass = "gr-qc",
    doi = "10.1103/PhysRevD.97.124037",
    journal = "Phys. Rev. D",
    volume = "97",
    number = "12",
    pages = "124037",
    year = "2018"
}

@article{Nielsen:2018lkf,
    author = "Nielsen, Alex B. and Capano, Collin D. and Birnholtz, Ofek and Westerweck, Julian",
    title = "{Parameter estimation and statistical significance of echoes following black hole signals in the first Advanced LIGO observing run}",
    eprint = "1811.04904",
    archivePrefix = "arXiv",
    primaryClass = "gr-qc",
    doi = "10.1103/PhysRevD.99.104012",
    journal = "Phys. Rev. D",
    volume = "99",
    number = "10",
    pages = "104012",
    year = "2019"
}

@article{Uchikata:2019frs,
    author = "Uchikata, Nami and Nakano, Hiroyuki and Narikawa, Tatsuya and Sago, Norichika and Tagoshi, Hideyuki and Tanaka, Takahiro",
    title = "{Searching for black hole echoes from the LIGO-Virgo Catalog GWTC-1}",
    eprint = "1906.00838",
    archivePrefix = "arXiv",
    primaryClass = "gr-qc",
    doi = "10.1103/PhysRevD.100.062006",
    journal = "Phys. Rev. D",
    volume = "100",
    number = "6",
    pages = "062006",
    year = "2019"
}

@article{Cardoso:2019rvt,
      author         = "Cardoso, Vitor and Pani, Paolo",
      title          = "{Testing the nature of dark compact objects: a status
                        report}",
      journal        = "Living Rev. Rel.",
      volume         = "22",
      year           = "2019",
      number         = "1",
      pages          = "4",
      doi            = "10.1007/s41114-019-0020-4",
      eprint         = "1904.05363",
      archivePrefix  = "arXiv",
      primaryClass   = "gr-qc",
      SLACcitation   = "%%CITATION = ARXIV:1904.05363;%%"
}

@article{Lu:2017vdx,
    author = "Lu, Wenbin and Kumar, Pawan and Narayan, Ramesh",
    title = "{Stellar disruption events support the existence of the black hole event horizon}",
    eprint = "1703.00023",
    archivePrefix = "arXiv",
    primaryClass = "astro-ph.HE",
    doi = "10.1093/mnras/stx542",
    journal = "Mon. Not. Roy. Astron. Soc.",
    volume = "468",
    number = "1",
    pages = "910--919",
    year = "2017"
}

@article{Doeleman:2012zc,
    author = "Doeleman, Sheperd S. and others",
    title = "{Jet Launching Structure Resolved Near the Supermassive Black Hole in M87}",
    eprint = "1210.6132",
    archivePrefix = "arXiv",
    primaryClass = "astro-ph.HE",
    doi = "10.1126/science.1224768",
    journal = "Science",
    volume = "338",
    pages = "355",
    year = "2012"
}

@article{EventHorizonTelescope:2019dse,
    author = "Akiyama, Kazunori and others",
    collaboration = "Event Horizon Telescope",
    title = "{First M87 Event Horizon Telescope Results. I. The Shadow of the Supermassive Black Hole}",
    eprint = "1906.11238",
    archivePrefix = "arXiv",
    primaryClass = "astro-ph.GA",
    doi = "10.3847/2041-8213/ab0ec7",
    journal = "Astrophys. J. Lett.",
    volume = "875",
    pages = "L1",
    year = "2019"
}

@article{EventHorizonTelescope:2019ggy,
    author = "Akiyama, Kazunori and others",
    collaboration = "Event Horizon Telescope",
    title = "{First M87 Event Horizon Telescope Results. VI. The Shadow and Mass of the Central Black Hole}",
    eprint = "1906.11243",
    archivePrefix = "arXiv",
    primaryClass = "astro-ph.GA",
    doi = "10.3847/2041-8213/ab1141",
    journal = "Astrophys. J. Lett.",
    volume = "875",
    number = "1",
    pages = "L6",
    year = "2019"
}

@article{Zulianello:2020cmx,
    author = "Zulianello, Anna and Carballo-Rubio, Ra\'ul and Liberati, Stefano and Ansoldi, Stefano",
    title = "{Electromagnetic tests of horizonless rotating black hole mimickers}",
    eprint = "2005.01837",
    archivePrefix = "arXiv",
    primaryClass = "gr-qc",
    doi = "10.1103/PhysRevD.103.064071",
    journal = "Phys. Rev. D",
    volume = "103",
    number = "6",
    pages = "064071",
    year = "2021"
}

@article{Carballo-Rubio:2018jzw,
    author = "Carballo-Rubio, Ra\'ul and Di Filippo, Francesco and Liberati, Stefano and Visser, Matt",
    title = "{Phenomenological aspects of black holes beyond general relativity}",
    eprint = "1809.08238",
    archivePrefix = "arXiv",
    primaryClass = "gr-qc",
    doi = "10.1103/PhysRevD.98.124009",
    journal = "Phys. Rev. D",
    volume = "98",
    number = "12",
    pages = "124009",
    year = "2018"
}

@article{Teukolsky:1973ha,
    author = "Teukolsky, Saul A.",
    title = "{Perturbations of a rotating black hole. 1. Fundamental equations for gravitational electromagnetic and neutrino field perturbations}",
    doi = "10.1086/152444",
    journal = "Astrophys. J.",
    volume = "185",
    pages = "635--647",
    year = "1973"
}

@article{Teukolsky:1972my,
      author         = "Teukolsky, S. A.",
      title          = "{Rotating black holes - separable wave equations for
                        gravitational and electromagnetic perturbations}",
      journal        = "Phys. Rev. Lett.",
      volume         = "29",
      year           = "1972",
      pages          = "1114-1118",
      doi            = "10.1103/PhysRevLett.29.1114",
      reportNumber   = "OAP-291",
      SLACcitation   = "%%CITATION = PRLTA,29,1114;%%"
}

@article{Mathur:2005zp,
    author = "Mathur, Samir D.",
    editor = "Kiritsis, E.",
    title = "{The Fuzzball proposal for black holes: An Elementary review}",
    eprint = "hep-th/0502050",
    archivePrefix = "arXiv",
    doi = "10.1002/prop.200410203",
    journal = "Fortsch. Phys.",
    volume = "53",
    pages = "793--827",
    year = "2005"
}

@article{Guo:2017jmi,
    author = "Guo, Bin and Hampton, Shaun and Mathur, Samir D.",
    title = "{Can we observe fuzzballs or firewalls?}",
    eprint = "1711.01617",
    archivePrefix = "arXiv",
    primaryClass = "hep-th",
    doi = "10.1007/JHEP07(2018)162",
    journal = "JHEP",
    volume = "07",
    pages = "162",
    year = "2018"
}

@article{Berti:2007zu,
    author = "Berti, Emanuele and Cardoso, Jaime and Cardoso, Vitor and Cavaglia, Marco",
    title = "{Matched-filtering and parameter estimation of ringdown waveforms}",
    eprint = "0707.1202",
    archivePrefix = "arXiv",
    primaryClass = "gr-qc",
    doi = "10.1103/PhysRevD.76.104044",
    journal = "Phys. Rev. D",
    volume = "76",
    pages = "104044",
    year = "2007"
}

@article{Berti:2005gp,
      author         = "Berti, Emanuele and Cardoso, Vitor and Casals, Marc",
      title          = "{Eigenvalues and eigenfunctions of spin-weighted
                        spheroidal harmonics in four and higher dimensions}",
      journal        = "Phys. Rev.",
      volume         = "D73",
      year           = "2006",
      pages          = "024013",
      doi            = "10.1103/PhysRevD.73.109902, 10.1103/PhysRevD.73.024013",
      note           = "[Erratum: Phys. Rev.D73,109902(2006)]",
      eprint         = "gr-qc/0511111",
      archivePrefix  = "arXiv",
      primaryClass   = "gr-qc",
      SLACcitation   = "%%CITATION = GR-QC/0511111;%%"
}

@online{pycbcgithub,
    author = "A.~H.~Nitz and others",
    title = "PyCBC Software",
    url = {https://github.com/gwastro/pycbc, GitHub},
    year = "2021"
}

@Article{Biwer:2018osg,
  author        = {Biwer, C. M. and Capano, Collin D. and De, Soumi and Cabero, Miriam and Brown, Duncan A. and Nitz, Alexander H. and Raymond, V.},
  title         = {{PyCBC Inference: A Python-based parameter estimation toolkit for compact binary coalescence signals}},
  journal       = {Publ. Astron. Soc. Pac.},
  year          = {2019},
  volume        = {131},
  number        = {996},
  pages         = {024503},
  archiveprefix = {arXiv},
  doi           = {10.1088/1538-3873/aaef0b},
  eprint        = {1807.10312},
  primaryclass  = {astro-ph.IM},
}

@Article{ForemanMackey:2012ig,
  author        = {Foreman-Mackey, Daniel and Hogg, David W. and Lang, Dustin and Goodman, Jonathan},
  title         = {{emcee: The MCMC Hammer}},
  journal       = {Publ. Astron. Soc. Pac.},
  year          = {2013},
  volume        = {125},
  pages         = {306--312},
  archiveprefix = {arXiv},
  doi           = {10.1086/670067},
  eprint        = {1202.3665},
  primaryclass  = {astro-ph.IM},
}

@Article{Vousden:2015,
  author   = {Vousden, W. D. and Farr, W. M. and Mandel, I.},
  title    = {{Dynamic temperature selection for parallel tempering in Markov chain Monte Carlo simulations}},
  journal  = {Monthly Notices of the Royal Astronomical Society},
  year     = {2015},
  volume   = {455},
  number   = {2},
  pages    = {1919-1937},
  month    = {11},
  issn     = {0035-8711},
  doi      = {10.1093/mnras/stv2422},
  eprint   = {https://academic.oup.com/mnras/article-pdf/455/2/1919/18514064/stv2422.pdf},
  url      = {https://doi.org/10.1093/mnras/stv2422},
}

@article{LIGOScientific:2021sio,
    author = "Abbott, R. and others",
    collaboration = "LIGO Scientific, VIRGO, KAGRA",
    title = "{Tests of General Relativity with GWTC-3}",
    eprint = "2112.06861",
    archivePrefix = "arXiv",
    primaryClass = "gr-qc",
    reportNumber = "LIGO-P2100275",
    month = "12",
    year = "2021"
}

@article{Westerweck:2021nue,
    author = "Westerweck, Julian and Sherf, Yotam and Capano, Collin D. and Brustein, Ram",
    title = "{Sub-atomic constraints on the Kerr geometry of GW150914}",
    eprint = "2108.08823",
    archivePrefix = "arXiv",
    primaryClass = "gr-qc",
    month = "8",
    year = "2021"
}

@article{Virtanen:2019joe,
    author = "Virtanen, Pauli and others",
    title = "{SciPy 1.0--Fundamental Algorithms for Scientific Computing in Python}",
    eprint = "1907.10121",
    archivePrefix = "arXiv",
    primaryClass = "cs.MS",
    doi = "10.1038/s41592-019-0686-2",
    journal = "Nature Meth.",
    volume = "17",
    pages = "261",
    year = "2020"
}

@article{Barausse:2018vdb,
    author = "Barausse, Enrico and Brito, Richard and Cardoso, Vitor and Dvorkin, Irina and Pani, Paolo",
    title = "{The stochastic gravitational-wave background in the absence of horizons}",
    eprint = "1805.08229",
    archivePrefix = "arXiv",
    primaryClass = "gr-qc",
    doi = "10.1088/1361-6382/aae1de",
    journal = "Class. Quant. Grav.",
    volume = "35",
    number = "20",
    pages = "20LT01",
    year = "2018"
}

@article{Ren:2021xbe,
    author = "Ren, Jing and Wu, Di",
    title = "{Gravitational wave echoes search with combs}",
    eprint = "2108.01820",
    archivePrefix = "arXiv",
    primaryClass = "gr-qc",
    doi = "10.1103/PhysRevD.104.124023",
    journal = "Phys. Rev. D",
    volume = "104",
    number = "12",
    pages = "124023",
    year = "2021"
}

@article{Wu:2023wfv,
    author = "Wu, Di and Gao, Pengyuan and Ren, Jing and Afshordi, Niayesh",
    title = "{Model-independent search for the quasinormal modes of gravitational wave echoes}",
    eprint = "2308.01017",
    archivePrefix = "arXiv",
    primaryClass = "gr-qc",
    doi = "10.1103/PhysRevD.108.124006",
    journal = "Phys. Rev. D",
    volume = "108",
    number = "12",
    pages = "124006",
    year = "2023"
}

@article{Vishveshwara:1970zz,
    author = "Vishveshwara, C. V.",
    title = "{Scattering of Gravitational Radiation by a Schwarzschild Black-hole}",
    doi = "10.1038/227936a0",
    journal = "Nature",
    volume = "227",
    pages = "936--938",
    year = "1970"
}

@article{Cornish:2014kda,
    author = "Cornish, Neil J. and Littenberg, Tyson B.",
    title = "{BayesWave: Bayesian Inference for Gravitational Wave Bursts and Instrument Glitches}",
    eprint = "1410.3835",
    archivePrefix = "arXiv",
    primaryClass = "gr-qc",
    doi = "10.1088/0264-9381/32/13/135012",
    journal = "Class. Quant. Grav.",
    volume = "32",
    number = "13",
    pages = "135012",
    year = "2015"
}

@article{Drago:2020kic,
    author = "Drago, M. and others",
    title = "{Coherent WaveBurst, a pipeline for unmodeled gravitational-wave data analysis}",
    journal = {SoftwareX},
    volume = {14},
    pages = {100678},
    eprint = "2006.12604",
    archivePrefix = "arXiv",
    primaryClass = "gr-qc",
    doi = "10.1016/j.softx.2021.100678",
    month = "6",
    year = "2020"
}

@article{Datta:2020gem,
    author = "Datta, Sayak and Phukon, Khun Sang and Bose, Sukanta",
    title = "{Recognizing black holes in gravitational-wave observations: Challenges in telling apart impostors in mass-gap binaries}",
    eprint = "2004.05974",
    archivePrefix = "arXiv",
    primaryClass = "gr-qc",
    reportNumber = "LIGO-P2000115",
    doi = "10.1103/PhysRevD.104.084006",
    journal = "Phys. Rev. D",
    volume = "104",
    number = "8",
    pages = "084006",
    year = "2021"
}

@article{Mukherjee:2022wws,
    author = "Mukherjee, Samanwaya and Datta, Sayak and Tiwari, Srishti and Phukon, Khun Sang and Bose, Sukanta",
    title = "{Toward establishing the presence or absence of horizons in coalescing binaries of compact objects by using their gravitational wave signals}",
    eprint = "2202.08661",
    archivePrefix = "arXiv",
    primaryClass = "gr-qc",
    reportNumber = "LIGO-P2100474",
    doi = "10.1103/PhysRevD.106.104032",
    journal = "Phys. Rev. D",
    volume = "106",
    number = "10",
    pages = "104032",
    year = "2022"
}
\end{document}